\documentclass[12pt,twoside]{article}
\usepackage[mathscr]{eucal}
\usepackage{amsmath,amsfonts,amssymb,amsthm}
\usepackage[hypertex]{hyperref}
\usepackage[matrix,arrow]{xy}
\usepackage{mathabx}

\baselineskip
\advance\textheight\topskip
\numberwithin{equation}{section} \makeatletter
\@addtoreset{equation}{section}
\def\be{\begin{equation}}
\def\ee{\end{equation}}
\def\ba{\begin{array}}
\def\ea{\end{array}}
\def\dps{\displaystyle}
\newcommand{\half}{\frac{1}{2}}

\newtheorem{prop}{Proposition}[section]
\newtheorem{lemma}[prop]{Lemma}

\newcommand{\bref}[1]{\textbf{\ref{#1}}}

\renewcommand{\d}{\partial}

\def\cD{\mathcal{D}}

\def\cJ{\mathcal{J}}

\def\cO{\mathcal{O}}

\begin{document}

\begin{flushright}
FIAN-TD-2012-10 \\
ESI-preprint 2374\\
\end{flushright}


\begin{center}

{\LARGE\textbf{
Mixed-symmetry tensor conserved currents
\\ [8pt]
and  AdS/CFT correspondence }}
\vspace{.9cm}

{\large Konstantin  Alkalaev}

\vspace{0.5cm}

\textit{I.E. Tamm Department of Theoretical Physics, \\P.N.
Lebedev Physical Institute,\\ Leninsky ave. 53, 119991 Moscow,
Russia}


\begin{abstract}
We present  the full list of conserved currents built of two
massless spinor fields in Minkowski space and their derivatives
multiplied by Clifford algebra elements. The currents have
particular mixed-symmetry type described  by Young diagrams with
one row and one column of arbitrary lengths and heights. Along
with Yukawa-like  totally antisymmetric currents the complete  set
of constructed currents exactly matches the spectrum of AdS
mixed-symmetry fields arising in the generalized Flato-Fr\o nsdal
theorem for  two spinor singletons. As a by-product, we formulate
and study general properties of primary fields  and conserved
currents of mixed-symmetry type.

\end{abstract}

\end{center}


\section{Introduction}
The $AdS_{d+1}/CFT_d$ correspondence establishes a dictionary between bulk gauge fields and boundary conserved currents
that in the case of totally symmetric massless fields of arbitrary spin is guaranteed by the generalized
Flato-Fr\o nsdal theorem for two scalar singletons \cite{Flato:1978qz,Vasiliev:2004cm}. Boundary
conserved currents describe then composite excitations of  two massless scalar fields (see, \textit{e.g.,}
\cite{Jevicki:2011ss} for reviews and more references).

It is remarkable that the above picture  extends beyond the case of totally symmetric bulk fields and boundary
currents to include tensors  of mixed-symmetry type on both sides of the correspondence.
The generalized Flato-Fr\o nsdal theorem plays  the central role once again but now it involves
two spinor singletons.
Indeed, as shown by Vasiliev   the tensor product of two $o(d,2)$ spinor singletons
decomposes into an infinite  direct sum of bosonic massless $AdS_{d+1}$ fields of mixed-symmetry "hook" type  and a
finite set of massive totally antisymmetric fields including massive scalar  \cite{Vasiliev:2004cm}.
\footnote{The Flato-Fr\o nsdal theorem has been originally established in $d=3$ dimensions, where
only totally symmetric $o(3,2)$ modules appear. Mixed-symmetry type $o(d,2)$ modules
arise in higher dimensions, $d\geq 4$.}

The problem we solve in this paper is to explicitly construct  conserved currents of mixed-symmetry type
that underlie boundary field-theoretical realization of the generalized Flato-Fr\o nsdal theorem.
The corresponding currents that we further refer to as Flato-Fr\o nsdal currents
are built of two massless spinors on Minkowski space and a finite number $(s-1)\geq 0$ of spacetime derivatives multiplied by
Clifford algebra elements of rank $p \leq [d/2]$. Their conformal dimensions given by
\be
\label{Dcrit}
\Delta_{crit} = s+d-2\;,
\ee
 saturate the unitary bound of $o(d,2)$ infinite-dimensional highest weight modules of mixed-symmetry
"hook" type \cite{Metsaev:1995re}. The currents are $o(d,2)$ primary fields of mixed-symmetry type
and their conservation conditions are conformally invariant, \textit{i.e.}, these are primary fields
as well.

The paper is organized as follows. We start  our analysis in section  \bref{sec:conf}
with discussion of conformal primary fields of mixed-symmetry type described
by Young diagrams with one row and one column.
Studying  conformal invariance in Section \bref{sec:prima}
we find  two types of conformally invariant conditions with one derivative  to be identified later on  with current conservation laws, and
fix corresponding conformal dimensions. In Section \bref{sec:twopoint} we
build conformally invariant two-point correlation functions. Making use of  the fact that any conformal
transformation  is a combination of Poincare transformations and inversions, we express two-point correlator
in terms of a single matrix proportional to Jacobi matrix of the inversion. More precisely,
the correlator is written in terms of two particular symmetry combinations of the matrix with symmetric and
antisymmetric indices. Section \bref{sec:confFF} considers advertised   mixed-symmetry conserved currents in CFT. In Section
\bref{sec:genformcurrent} we develop a general theory of conserved mixed-symmetry currents,
paying particular attention to trivially conserved currents, the so-called "improvements". In the following
sections we give particular realization of mixed-symmetry primary fields
and build conserved currents of  scale dimensions and symmetry types that exactly match the  Flato-Fr\o nsdal
spectrum. So, in Section \bref{sec:gen} a convenient bi-local generating formulation
is considered that produces conserved currents as expansion coefficients in auxiliary commuting and anticommuting
variables. It is worth noting that using bi-local functions makes sense from the holographic perspective (see, \textit{e.g.}, the second item  in
\cite{Jevicki:2011ss}). In Section \bref{sec:FF}, using the generating formulation  we build traceful
Flato-Fronsdal conserved currents. In Section \bref{sec:exa} we consider an example of building a traceless
conserved current described by three cell hook Young diagram. Conclusions and outlooks are collected  in Section \bref{sec:conclu}.

In the rest of the introduction section  we briefly review  $o(d,2)$ infinite-dimensional modules identified with
mixed-symmetry "hook" massless fields in $AdS_{d+1}$ and  their realization  via  the generalized Flato-Fr\o nsdal theorem
in the sector of two spinor singletons.

\subsection{Review of  the generalized Flato-Fr\o nsdal theorem }
\label{sec:FF}

We consider $AdS_{d+1}$ bosonic massless fields with particular spins organized  as  Young diagrams with one row of length $s$ and one column of
height $p$. \footnote{Field-theoretical description of $AdS_{d+1}$ mixed-symmetry free fields is available
now within various approaches
\cite{Metsaev:1995re,Brink:2000ag,Metsaev:2002vr, Zinoviev:2002ye,Alkalaev:2006rw,Boulanger:2008kw,
Skvortsov:2009zu,Alkalaev:2011zv,Burdik:2011cb,Campoleoni:2012th}. Also, some cubic interaction vertices between
mixed-symmetry AdS fields and the gravity are known
\cite{Alkalaev:2010af}.} Corresponding  unitary infinite-dimensional highest weight modules of $o(d,2)$ algebra
are denoted as $\cD(E_0|\, s,p)$, where $o(2)$ weight is the energy $E_0$ is   computed by Metsaev  \cite{Metsaev:1995re}
\be
\label{energsym}
E_0 = s+d-2\;,
\ee
and $o(d)$ spin weights  $(s,\overbrace{1,1,...,1,}^{p-1} 0,..., 0)$, where $s> 1,\; p \leq [d/2]$,  are depicted as a couple of numbers $s,p$.
In the sequel, tensors of such "hook" symmetry type are referred to as $\{ s, p \}$-type tensors.
The case $s=1, p\geq 1$
of totally antisymmetric massless fields is special, the respective energy is given by formula \eqref{energsymNU} below.
Note  that the same value of energy holds for totally symmetric
massless fields $\cD(s+d-2|\, s)$ because $E_0$ for unitary fields of general mixed-symmetry type depends on the size of the uppermost
rectangular block only \cite{Metsaev:1995re}.
For instance, a simplest "hook" $s=2$, $p=2$ field has  the energy $ E_0 = d$,
which value coincides with that of the $AdS_{d+1}$ graviton field.

Non-unitary  massless modules $\cD(E_0|\, s,p)$ of the same symmetry type have values of the energy
given by \cite{Metsaev:1995re} \footnote{Massless fields with energies below the unitary limit \eqref{Dcrit}
 = \eqref{energsym}
correspond to non-unitary $o(d,2)$ modules. Their gauge symmetries have different structure compared to that of
unitary massless fields, see \cite{Metsaev:1995re, Brink:2000ag}. For instance, a unitary  "hook" field
$s=2,\, p=2$ has the only gauge parameter given by antisymmetric Lorentz tensor, while its non-unitary
cousin has the only gauge parameter given by symmetric Lorentz tensor. On the other hand, spin $s=2,\, p=2$
massless field in Minkowski space has both gauge symmetries.}
\be
\label{energsymNU}
E_0 = d-p\;.
\ee
Note  that the same value of energy holds for unitary totally antisymmetric
massless fields. As a consequence, values of $E_0$  for height $p$ "hook" non-unitary fields  and
$p$-form massless unitary fields coincide.
For instance, a simplest "hook" $s=2$, $p=2$ field has the energy $ E_0 = d-2$,
which value coincides with that of the $AdS_{d+1}$ Kalb-Ramond field.

The remarkable fact about unitary mixed-symmetry
modules $\cD(E_0|\, s,p)$ of arbitrary spins is that all of them appear in the tensor product of two
Dirac spinor singletons. This statement is
the content of the Flato-Fr\o nsdal theorem  that has been
originally established for $o(3,2)$ algebra  (in this case
$p=1$, so only totally symmetric representations appear)
\cite{Flato:1978qz}. More recently, the theorem has been
extended by Vasiliev to the case of $o(d,2)$ algebras with any $d$ and it has been
shown that the whole scope of unitary "hook"  representations with
all admissible values of spins $s$ and  $p$ appears \cite{Vasiliev:2004cm}.
\footnote{In $d=4$ case $o(4,2)$ modules of arbitrary mixed-symmetry type  arise in the tensor product of two
(super)doubletons \cite{Gunaydin:1984fk}.}

To formulate the theorem we consider   Dirac spinor singleton
 defined as $o(d,2)$ infinite-dimensional highest weight unitary module with quantum numbers
\be
\label{singleton}
E_0 = \frac{d-1}{2}\;,
\qquad
\text{and}
\qquad
s=(\half, 0, ..., 0)\;,
\ee
and introduce notation
${\rm Di}=\cD(\frac{d-1}{2}, \half)$. According to Ref. \cite{Vasiliev:2004cm},
the tensor product of two spinor singletons decomposes into a direct sum
of mixed-symmetry "hook" $o(d,2)$ modules discussed above
\be
\label{FF}
\ba{c}
\dps
{\rm Di} \otimes {\rm Di}
=\cD(d-1,0)\oplus \sum_{s=1}^\infty \oplus \sum_{p=1}^{[d/2]}\oplus \,\Big(\cD(s+d-2 \,|\,s, p)\oplus
\\
\\
\dps
\hspace{20mm}\oplus \cD_{+}(s+d-2 \,|\, s, d/2)\oplus \cD_{-}(s+d-2 \,|\, s, d/2)\Big)\;,
\ea
\ee
where (anti-)selfdual modules $\cD_{\pm}(s+d-2 \,|\, s, d/2)$ appear in even dimension only. Note that
modules $\cD(d-1,0)$ and $\cD(d-1 \,|\, 1, m)$, $m=2,3,...$ correspond to unitary  scalar and antisymmetric fields which are massive  because their energies
are above the critical values \eqref{energsymNU}. All  other fields
in the above decomposition are unitary massless fields.

\vspace{2mm}

In the following sections we develop a realization of the generalized Flato-Fronsdal theorem  \eqref{FF} via conserved
currents built of conformal spinor fields. To this end we first  describe non-symmetric $o(d,2)$ primary fields
and respective conformally-invariant conditions to be identified with currents and their conservation laws.

\section{Conformal primary fields of mixed-symmetry type }
\label{sec:conf}

We consider the conformal algebra  $o(d,2)$ acting on  $d$-dimensional Minkowski spacetime $\mathbb{R}^{d-1,1}$ in a standard
fashion by coordinate transformations leaving the interval  $ds^2 = \eta_{mn} dx^m dx^n$ invariant up to
an  $x$-dependent positive scale factor, $m,n = 0,...,d-1$. The respective Killing vector can be represented in the form
\be
\label{killing}
\xi_m(x) = P_m + \Lambda_{mn}x^n + D x_m + K_m x^2 -2 x_m (Kx)\;,
\ee
where $P_m$, $\Lambda_{mn} = - \Lambda_{nm}$, $D$, and $K_m$ are constant $o(d-1,1)$ tensors that
parameterize elements of the conformal group under consideration.

One of the main goals of the present section is to formulate conformally invariant conditions that can be imposed
on a given mixed-symmetry primary field. This classification problem has been exhaustively solved in Ref. \cite{Shaynkman:2004vu}
by using a powerful technique based on the unfolded formulation originally elaborated to describe higher spin dynamics.
Conformal operators and invariant conditions for $SU^*(4)$ group  were  also studied  in Ref. \cite{DobrevPetkova}, while
$O(4,2)$ conformal operators realized as basis elements of an OPE decomposition for matter  fields in $\mathbb{R}^{3,1}$
were discussed in Refs. \cite{Dobrev}.
However, for our present purposes of considering particular symmetry type
primary fields  more efficient and immediate way is to use the standard field-theoretical
construction elaborated for lower-spin cases (see, \textit{e.g.}, \cite{Osborn:1993cr}).

The main idea behind the search for conformally invariant equations is to study secondaries for a given primary state
and find out when  the  secondary becomes  primary again so that conformal invariance remains unbroken.
From a group-theoretical perspective, such states are singular vectors in  modules generated from
primary states. In the conformal field theory under consideration primary states and their singular secondaries are   realized as currents and
conservation conditions.

\subsection{Primary fields and  conformally invariant conditions}
\label{sec:prima}

Conformal primary  field $\cO^{A}(x)$, where $x\in \mathbb{R}^{d-1,1}$ and $A$ collectively denotes components of $o(d-1,1)$ irreducible representation
transforms with respect to conformal Killing vector \eqref{killing} as \cite{Mack:1969rr}
\be
\label{prim}
\delta_\xi \cO^A(x) = - \big(L^{\Delta}_\xi \cO\big)^A(x)\;,
\qquad
L^\Delta_\xi = \xi^m\d_m + \Delta \sigma_\xi  - \frac{1}{4}(\d_m \xi_n  - \d_n \xi_m) \Sigma^{mn}\;,
\ee
where   $\big(\Sigma^{mn}\big)^A{}_B = - \big(\Sigma^{nm}\big)^A{}_B $ are $o(d-1,1)$ matrices, derivative $\d_m = \d/\d x^m$, parameter $\sigma_\xi$
is a combination $\sigma_\xi = D - 2 (Kx)$, and $\Delta$ is a
conformal dimension.

Let us consider particular primary field  which is a traceless rank-3 $o(d-1,1)$ tensor $\cO_{m|n|k}(x)$,
where slashes mean that indices are not related to each other by any type of permutation symmetry. Its transformation law \eqref{prim} is given by
\be
\ba{r}
\dps
\delta_\xi \cO_{m|n|k}(x) = - \big(\xi^l\d_l + \Delta \sigma_\xi )\cO_{m|n|k}(x)
\\
\\
\dps
+ \big(\Lambda_{mp}+2(K_m x_p - K_p x_m)\big) \cO^{p}{}_{|n|k}(x)
\\
\\
\dps
+ \big(\Lambda_{np}+2(K_n x_p - K_p x_n)\big) \cO_{m|}{}^{p}{}_{|k}(x)
\\
\\
\dps
+ \big(\Lambda_{kp}+2(K_k x_p - K_p x_k)\big) \cO_{m|n|}{}^{p}(x)\;,
\\

\ea
\ee
where we used $o(d-1,1)$ rank-3   representation matrices,
\be
\big(\Sigma^{mn}\big)_{\alpha i| \beta j|\gamma k} =
\big(\Sigma^{mn}\big)_{\alpha i} \eta_{\beta j}\eta_{\gamma k}+
\big(\Sigma^{mn}\big)_{\beta j} \eta_{\alpha i}\eta_{\gamma k}+
\big(\Sigma^{mn}\big)_{\gamma k} \eta_{\alpha i}\eta_{\beta j}\;,
\ee
and $\big(\Sigma^{mn}\big)_{\alpha i}  = \delta^m_\alpha \delta^n_i -\delta^n_\alpha \delta^m_i$
are $o(d-1,1)$ vector representation matrices, $\alpha, \beta, \gamma, i,j, k = 0,..., d-1$.

Obviously, taking the divergences on the primary field gives rise to additional inhomogeneous terms
in the transformation law. For the case at hand, one obtains
\be
\label{diverprim}
\ba{l}
\dps
\delta_\xi \d^m \cO_{m|n|k}(x) =  - L_\xi^{\Delta+1}\big(\d^m \cO_{m|n|k}(x)\big)
\\
\\
\dps
\hspace{32mm} +2K^\alpha\Big((\Delta-d+1)
\cO_{\alpha|n|k}(x)  - \cO_{n|\alpha|k}(x) - \cO_{k|n|\alpha}(x)\Big)\;.
\ea
\ee
As indicated in the first term above, a conformal weight of the
secondary field  $\d^m \cO_{m|n|k}(x)$ increases  by one compared
to that of the primary field. Generally, inhomogeneous terms in
the second line are non-vanishing, but for particular symmetry
type of indices these may cancel that results in  special values
of conformal weight $\Delta$. Traceless tensor $\cO_{m|n|k}(x)$
can be decomposed into three irreducible components according to
their symmetry type. Namely, these are a totally symmetric
traceless component $\cO_{(mnk)}(x)$, a totally antisymmetric
component $\cO_{[mnk]}(x)$, and a "hook"  traceless  component
$\cO_{mn,k}(x)$ symmetric in first two indices and satisfying
Young symmetry condition  $\cO_{(mn,k)}(x) \equiv 0$.

For totally symmetric and antisymmetric components we conclude that secondary fields
transform homogeneously provided dimensions take values $\Delta = d+1$ and $\Delta = d-3$, which
are  recognized as critical conformal weights, cf. \eqref{energsym}, \eqref{energsymNU}.
It means precisely that
one may impose conditions  $\d^m \cO_{(mnk)}(x) =0$ or $\d^m \cO_{[mnk]}(x) =0$ that define invariant
subspaces in modules generated from the corresponding primary fields.

For the "hook" component $\cO_{mn,k}(x)$ the situation is slightly more subtle. According to general formula
\eqref{diverprim} secondary field  $\d^m \cO_{mn,k}(x)$ does not transform homogeneously and only its
symmetric or antisymmetric parts do. For special  values of conformal dimension $\Delta$ one can impose
just one of the following constraints at once, \footnote{From now on both symmetrization and antisymmetrization
denoted respectively by $(...)$ and $[...]$ come with a unit weight. }
\be
\label{dimesnions}
\ba{c}
\dps
\;\;\;\;\Delta = d\;: \qquad \;\;\;\d^m \cO_{m[n,k]}(x) =0\;,
\\
\\
\dps
\Delta = d-2\;: \qquad \d^m \cO_{m(n,k)}(x) =0\;.
\\

\ea
\ee
Note that the last constraint is equivalent to $\d^m \cO_{nk,m}(x) =0$ by virtue of Young symmetry property.
From the $AdS_{d+1}/CFT_d$ correspondence perspective the first value of conformal weight in \eqref{dimesnions}
equals to  energy $E_0 = d$  of a unitary "hook" field in $AdS_{d+1}$, see \eqref{energsym},
while the second one equals to the energy $E_0 = d-2$ of a non-unitary "hook" field, see \eqref{energsymNU}.

The above discussion can be easily generalized to the case of  primary fields of more general symmetry type
$\cO_{m_1...m_s, n_1 ... n_{p-1}}(x)$, where the first group of indices is totally symmetric,
the second group is totally antisymmetric, while their mixed symmetry is described  by permutation condition
corresponding to $\{s,p\}$-type  Young diagram,
\be
\label{YTO}
\cO_{m_1... m_s, \, n_1 ... n_{p-1}}(x)\;:\qquad \cO_{(m_1... m_s, \, m_{s+1}) n_2 ... n_{p-1}}(x) \equiv 0\;,
\ee
along with the tracelessness condition,
\be
\;\;\qquad \qquad \qquad \qquad \qquad \eta^{nk}\cO_{nk m_3... m_s, \, n_1 ... n_{p-1}}(x) = 0\;.
\ee
Note that taking a cross-trace yields zero by virtue of Young  symmetry \eqref{YTO}.

The analysis of secondary field conformal transformations reveals  the following
conformally invariant conditions and conformal weights
\be
\label{dimesnions2}
\Delta = s+d-2\;: \quad \;\d^{k} \cO_{k a_2...a_{s-1}a_{s}, \,m_1m_2... m_{p-1}}(x)  -
\d^{k} \cO_{k (a_2...a_{s-1}[m_1,a_s) \,m_2... m_{p-1}]}(x) =0\;,
\ee
\be
\label{dimesnions3}
\Delta = d-p\;: \qquad \d^k \cO_{m_1...m_s, n_1 ... n_{p-2}k}(x) =0\;,
\ee
where particular symmetrizations used above correspond to $\{s-1,p\}$-type and $\{s, p-1\}$-type Young diagrams.

It is worth noting that conformally invariant condition \eqref{dimesnions2} is valid provided that conformal
weight takes its critical value \eqref{Dcrit}, or  \eqref{energsym}; conformally invariant condition
\eqref{dimesnions3} is valid provided that the conformal weight is equal to energy \eqref{energsymNU}.

\subsection{Two-point correlation functions}
\label{sec:twopoint}

Conformal invariance guarantees that two-point correlation
function of two
primary fields of coinciding dimensions $\Delta$ is completely fixed modulo a  prefactor. The general analysis
of Ref. \cite{Osborn:1993cr} shows that the form of two-point correlator can be concisely described
in terms of matrix
\be
\label{invma}
I_{mn} = \eta_{mn} - 2\frac{x_m x_n}{x^2}\;,
\ee
which is proportional to Jacobi matrix for the inversion transformation $x^i \rightarrow x^i/x^2$. The construction
heavily relies upon the fact that any conformal transformations can be generated by
Poincare transformations and inversions. Then, the answer for two primary operators of $\{s,p\}$ mixed-symmetry type
\eqref{YTO} is given in terms of two (anti)symmetric tensors,
\be
\label{buildblocks}
\ba{l}
\dps
S_{i_1 ... i_s| j_1 ... j_s} = I_{(i_1 j_1} \cdots I_{i_s) j_s}\;,
\qquad\;
A_{m_1 ... m_{p-1}| n_1 ... n_{p-1}} = I_{[m_1 n_1} \cdots I_{m_{p-1}] n_{p-1}}\;,
\dps
\ea
\ee
where (anti)symmetrizations are performed with respect to the first groups of indices, while the second groups of
indices are automatically (anti)symmetrized. Therefore,
$S_{i_1 ... i_s| j_1 ... j_s} = S_{(i_1 ... i_s)| (j_1 ... j_s)}$
and
$A_{m_1 ... m_{p-1}| n_1 ... n_{p-1}} = A_{[m_1 ... m_{p-1}]| [n_1 ... n_{p-1}]}$. These  tensors are building blocks
for the searched-for two-point correlation functions. Up to terms with traces, the answer is given by
\be
\label{2pointG}
\langle
\cO_{i_1...i_s,\,m_1 ... m_{p-1}}(x)\, \cO_{j_1 ... j_s,\,n_1 ... n_{p-1}}(0)\rangle  = C\,
\frac{\Pi_{i_1...i_s,\,m_1 ... m_{p-1}| j_1 ... j_s,\,n_1 ... n_{p-1}}(x)}{|x|^{2\Delta}}\;,
\ee
where $C$ is a normalization, $\Delta$ is a conformal weight, and projector
$\Pi_{...|...}(x)$ in the numerator is realized in terms of building blocks of ranks $s-1$ and $p$, respectively,
\be
\label{coooc}
\ba{l}
\dps
\Pi_{i_1...i_s,\,m_1 ... m_{p-1}| j_1 ... j_s,\,n_1 ... n_{p-1}} =
S_{(i_1 ... i_{s-1}| (j_1 ... j_{s-1}}A_{i_s) m_1 ... m_{p-1}| j_s)n_1 ... n_{p-1}} \;+ \;\;\eta\text{-terms}\;.
\ea
\ee
Note that the projector admits another realization  in terms of building blocks of ranks $s$ and $p-1$
by way of a different symmetrization. However,  blocks \eqref{buildblocks} are composed in terms
of symmetric matrix \eqref{invma} so both realizations coincide up to a normalization constant.
A trace contribution can be uniquely fixed by adjusting Young symmetry and trace properties of both sides.
\footnote{Let us comment that the holographic  effective action depends
on traceless shadow fields, while two-point correlators appear as the second variation. It implies that though the first contribution in \eqref{coooc} is directly  obtained, the
trace terms are reconstructed by adjusting Young symmetry and
trace properties of the shadow fields and the correlator. In the presence of
the conformal anomaly,  trace contributions are to be considered more carefully, see, \textit{e.g.}, \cite{Henningson:1998gx}.}
The correlation function can be represented in a standard fashion with explicit $x$-dependence as a linear combination
 of terms $\dps \eta_{b_1b_2}...\eta_{c_1c_2}\frac{x^{a_1} \cdots x^{a_{2n}}}{|x|^{2\Delta+2n}}$.

As an explicit example we give two-point function for primary conformal fields of simplest "hook" symmetry type read off
from general formula \eqref{2pointG},
\be
\label{2point}
\langle
\cO_{ij,\,k}(x) \cO_{mn,\,l}(0)\rangle  = 2C\,
\frac{\Pi_{ij,\,k| mn,\,l}(x)}{|x|^{2\Delta}}\;,
\ee
where $C$  is a normalization, and projector $\Pi_{ij,\,k| mn,\,l}(x)$ has the form
\be
\label{2point1}
\ba{l}
\dps
\Pi_{ij,\,k| mn,\,l} =
\big(I_{im}I_{jn}+I_{in}I_{jm}\big)I_{kl}
\\
\\
\dps
\hspace{18mm}
-\half \big(I_{im}I_{kn}+I_{in}I_{km}\big)I_{jl}
-\half \big(I_{jm}I_{kn}+I_{jn}I_{km}\big)I_{il}
-\frac{2}{d-1}T_{ij,\,k| mn,\,l}\;,
\ea
\ee
where trace part $T_{ij,\,k| mn,\,l}$ is given by
\be
\ba{c}
\dps
T_{ij,\,k| mn,\,l} = \eta_{ij}\eta_{mn}I_{kl}
-\half \eta_{mn}\big(\eta_{ik}I_{jl}+\eta_{jk}I_{il}\big)
\\
\\
\dps
\hspace{15mm} -\half \eta_{ij}\eta_{ml}I_{kn}
+\frac{1}{4} \eta_{ml}\big(\eta_{ik}I_{jn}+\eta_{jk}I_{in}\big)
\\
\\
\dps
\hspace{17mm}-\half \eta_{ij}\eta_{nl}I_{km}
+\frac{1}{4} \eta_{nl}\big(\eta_{ik}I_{jm}+\eta_{jk}I_{im}\big)\;.
\ea
\ee
An equivalent form with explicit $x$-dependence involves 60 terms.

\section{Mixed-symmetry type conserved currents}

\label{sec:confFF}

We assume that currents under consideration are composed  of elementary  fields in $\mathbb{R}^{d-1,1}$
and their derivatives. If constituent fields are chosen to be conformal  (\textit{e.g.}, these are massless matter
fields considered  in the present paper), then the resulting current may be a primary
field of definite conformal (scale) dimension.
As discussed in the previous section, currents with special  conformal dimensions  may satisfy conservation laws that
correspond to conformally-invariant conditions, namely \eqref{dimesnions2} or
\eqref{dimesnions3}. For arbitrary  conformal dimensions  conservation conditions may still hold
but special conformal invariance breaks down, leaving intact invariance with respect to Poincare and scale
transformations only, cf. \eqref{diverprim}.
In the general case of non-conformal constituent fields (\textit{e.g.}, these are massive matter fields) it is
still possible to impose  one or two types of conservation conditions but underlying conformal invariance is obviously missing.

Note that totally symmetric conserved currents of any spin  composed of massless matter fields in in $\mathbb{R}^{d-1,1}$
were  built in the context of higher spin gauge theory in  Refs. \cite{Berends:1985xx,Anselmi:1999bb}.
For more discussion of totally symmetric currents on maximally symmetric spacetimes see, \textit{e.g.},
\cite{Prokushkin:1999xq,Vasiliev:1999ba,Gelfond:2003vh,Manvelyan:2004mb,Bekaert:2009ud}.
Conformal operators in $\mathbb{R}^{3,1}$ with  $p=2$
were considered  in Ref.  \cite{DobrevGanchev}, though an explicit check of conservation conditions
has not been demonstrated therein  (see, however,  \cite{DobrevPetkova,Dobrev,DobrevGanchev} for more discussion).

\subsection{General formulation}
\label{sec:genformcurrent}
Let us consider tensor currents $J_{a_1... a_s, \, m_1 ... m_{p-1}}(x)$, where $x\in \mathbb{R}^{d-1,1}$, $a,m = 0,...,d-1$.
The first group of indices is totally symmetric, the second group is totally antisymmetric,
while their mixed symmetry is described  by Young condition
\be
\label{YT}
J_{a_1... a_s, \, m_1 ... m_{p-1}}(x)\;:\qquad J_{(a_1... a_s, \, a_{s+1}) m_2 ... m_{p-1}}(x) \equiv 0\;,
\ee
\textit{i.e.}, the current has $\{s,p\}$ symmetry type, cf. \eqref{YTO}. Imposing tracelessness condition is optional.

In order to formulate the first type conservation condition
it is convenient to introduce auxiliary constant tensor  $\xi_1^{a_1...a_{s-1}, m_1 ... m_{p-1}}$ of $\{s-1,p\}$-type
so that currents
\be
\label{curcur}
J_n(x|\xi_1) = \xi_1^{a_2...a_{s-1}, m_1 ... m_{p-1}} J_{n a_2... a_s, \, m_1 ... m_{p-1}}(x)
\ee
satisfy conservation law
\be
\label{curcon}
\d^n J_n(x|\xi_1) = 0\;,
\qquad
\d^n = \frac{\d}{\d x_n}\;.
\ee
Demanding constant parameters $\xi_1^{a_1...a_{s-1}, m_1 ... m_{p-1}}$ to be traceless with respect to Minkowski
metric $\eta_{ab}$ makes  current \eqref{curcur} double traceless, while the conservation condition
\eqref{curcon} remains intact. Unless otherwise stated, parameter $\xi_1$ is assumed to be traceful. From \eqref{curcon} it follows that  currents are defined up to "improvements",
\textit{i.e.,} trivially conserved terms,
\be
\label{impr}
J_n(x|\xi_1) \sim J_n(x|\xi_1) + \d^m I_{nm}(x|\xi_1)\;,
\ee
where $I_{nm}(x|\xi_1)$ are local (finite number of derivatives of composite fields) antisymmetric tensor, $I_{nm} = - I_{mn}$.
\footnote{It is interesting to note that conserved currents  in $AdS$ spacetime can also be treated as
improvements in the class of pseudolocal expansions in  inverse powers of the cosmological constant
\cite{Prokushkin:1999xq}. Apparently, for massless constituent fields in Minkowski space this phenomenon
is unparalleled, though massive theories may exhibit analogous behavior.}
Note that improvements for   a given current $J_n(x|\xi_1)$ with a scale dimension $\Delta$
 have scale dimension $\Delta-1$.

Conserved currents \eqref{curcur}, \eqref{curcon} define global symmetries with mixed-symmetry constant parameters
$\xi_1$, generated by  conserved charges built in a standard fashion as
\be
\label{charge}
Q(\xi_1) = \int_{\mathbb{R}^{d-1}}  J^* (x|\xi_1)\;,
\ee
where  the star label  denotes $(d-1)$-from dual to the original current,
\be
\ba{c}
\dps
J^* (x|\xi_1)  = J^*_{n_1 ... n_{d-1}} (x|\xi_1)\,dx^{n_1}\wedge ... \wedge dx^{n_{d-1}}\;,
\\
\\
J^*_{n_1 ... n_{d-1}} (x|\xi_1) = \epsilon_{n_1 ... n_{d-1}m}J^m(x|\xi_1)\;.
\ea
\ee
Provided the original current is conserved, the dual form is obviously closed
\be
\label{closed}
d J^* (x|\xi_1) = 0\;,
\qquad
d = dx^n \frac{\d}{\d x^n}\;,
\ee
that defines the current cohomology. Trivial cohomology class is given by zero charges that are  produced by improvement currents.

To formulate the second type conservation condition one  introduces auxiliary constant parameter $\xi_2^{a_1...a_{s}, m_1 ... m_{p-2}}$ of
$\{s,p-1\}$ type so that currents
\be
\label{curcur2}
J_n(x|\xi_2) = \xi_2^{a_1...a_{s}, m_1 ... m_{p-2}} J_{a_1... a_s, \, m_1 ... m_{p-2} n}(x)
\ee
satisfy conservation law
\be
\label{curcon2}
\d^n J_n(x|\xi_2) = 0\;,
\qquad
\d^n = \frac{\d}{\d x_n}\;.
\ee
Demanding constant parameters $\xi_2^{a_1...a_{s-1}, m_1 ... m_{p-2}}$ to be traceless with respect to Minkowski
metric $\eta_{ab}$ makes  current \eqref{curcur2} double traceless, while the conservation condition
\eqref{curcon2} remains intact. Quite analogously to \eqref{charge} - \eqref{closed}, one introduces dual currents
$J^*(x|\xi_2)$ and conserved charges $Q(\xi_2)$. It follows that  that original  currents $J_n(x|\xi_2)$
satisfy the following equivalence relation
\be
\label{impr2}
J_n(x|\xi_2) \sim J_n(x|\xi_2) + \d^m K_{nm}(x|\xi_2)\;,
\ee
where $K_{nm}(x|\xi_2)$ are local (finite number of derivatives of composite fields) antisymmetric tensor,
$K_{nm} = - K_{mn}$.

A given mixed-symmetry current \eqref{YT} may satisfy either  or both types of conservation conditions.
Using one or another conservation law makes no essential complications compared  to totally symmetric case.
Therefore, it is useful to study  the current cohomology when both equivalence relations \eqref{impr} and \eqref{impr2}
are imposed. In particular, we describe the trivial cohomology class. To this end we introduce the following quantities,
\be
\label{Ps}
\ba{c}
\dps
P_{k|n}(x|\xi_1) = \xi_1^{a_1... a_{s-1}, m_1 ... m_{p-1}}P_{k a_1... a_{s-1}, m_1 ... m_{p-1}n}(x)\;,
\\
\\
P_{k|n}(x|\xi_2) = \xi_2^{a_1... a_s, m_1 ... m_{p-2}}P_{a_1 ... a_s, m_1 ... m_{p-2}kn}(x)\;,
\ea
\ee
and
\be
\label{P3}
P_n(x|\xi_3) = \xi_3^{a_2... a_s, m_1 ... m_{p}}P_{n a_2... a_s, m_1 ... m_{p}}(x)\;.
\ee
Here  $P_{a_1... a_s, m_1 ... m_{p}}(x)$ is some $\{s,p+1\}$-type  traceful tensor, and parameters
$\xi_1$, $\xi_2$, and $\xi_3$ are constant tensors of types $\{s-1,p\}$, $\{s,p-1\}$, and $\{s-1,p+1\}$, respectively.
Note that $P_{k|n}(x|\xi_2)  = - P_{n|k}(x|\xi_2)$, while $P_{k|n}(x|\xi_1)$ has no definite symmetry
type with respect to indices $k$ and $n$.

\begin{lemma}
\label{lemma0}
Let  a given  $\{s,p\}$-type current satisfy both conservation conditions \eqref{curcon}
and \eqref{curcon2}. The current represents the  trivial current cohomology class  iff
it has the form $J_n(x|\xi_1) = \d^m I_{nm}(x|\xi_1)$ and
$J_n(x|\xi_2) = \d^m K_{nm}(x|\xi_2)$, with
\be
\label{trivclass1}
\d^m I_{nm}(x|\xi_1) = \d^k P_{k|n}(x|\xi_1)
- \d^k P_{n|k}(x|\xi_1)\;,
\qquad
\d^m K_{nm}(x|\xi_2) = \d^kP_{k|n}(x|\xi_2)\;,
\ee
where $P_{k|n}(x|\xi_1)$ and $P_{k|n}(x|\xi_2)$ are defined by relations \eqref{Ps}, while
$P_n(x|\xi_3)$ \eqref{P3} satisfies
\be
\label{newcorv}
\d^n P_{n}(x|\xi_3)= 0\;.
\ee
\end{lemma}
\vspace{5mm}
\noindent  To prove the lemma one notice that all $gl(d)$ irreducible
components contained in $I_{nm}(x|\xi_1)$ and $K_{nm}(x|\xi_2)$ are in the
tensor products of an antisymmetric 2-tensor and parameters $\xi_1$ and $\xi_2$, respectively.
Then, one observes that the only common component is given by $\{s,p+1\}$  type tensor denoted
by $P_{a_1... a_s, m_1 ... m_{p}}(x)$ above. Formulas \eqref{trivclass1} describe then two possible
$\{s,p\}$-type projections of $P_{a_1... a_s, m_1 ... m_{p}}(x)$ contracted with one derivative.
Constraint   \eqref{newcorv} is a consistency condition for \eqref{trivclass1}.

It is worth noting that due to \eqref{newcorv} quantity $P_{n}(x|\xi_3)$ can be treated  as a conserved current.
This observation allows one to reformulate Lemma \bref{lemma0} in the following useful way:
any $\{s,p+1\}$-type current with the first type
conservation law defines a trivial cohomology class of $\{s,p\}$-type currents subject to conservation laws of both types.

\subsection{Generating function in auxiliary variables}
\label{sec:gen}

In what follows, we discuss  conserved currents built of massless spinor fields in $d$-dimensional Minkowski space.
To this end,  let us consider Dirac spinor field $\psi_\alpha(x)$, $\alpha = 1,..., 2^{[d/2]}$
on   spacetime $\mathbb{R}^{d-1,1}$.
 Clifford algebra is defined a set of antisymmetrized
combinations
\be
\label{scliff}
\Big\{\Gamma_{m_1... m_k} = \gamma_{m_{[1}}... \gamma_{m_k]}\;, \quad k =0,1,..., d \Big \}
\ee
of  matrices $(\gamma_m)^\alpha{}_\beta$ satisfying anticommutation relations $\gamma_m \gamma_n
+\gamma_n \gamma_m = 2\eta_{mn} \mathbb{I}_d$, where $\eta_{mn}$ is Minkowski metric.
By definition,  $k=0$ gives the unity matrix $\mathbb{I}_d$.
Dirac conjugate is defined as $\bar\psi  = \psi^\dagger \gamma_0$,
where $\dagger$ denotes a hermitian conjugation operation. The spinor field fulfills
Dirac massless equation $(\gamma_m)^\beta{}_\alpha\, \d^m \psi_\beta(x)=0$ and
$\d^m\bar\psi^\beta(x)(\gamma_m)^\alpha{}_\beta  = 0$ for the conjugated field.

It is convenient to describe mixed-symmetry tensor currents as expansion coefficients of some
generating function with respect to auxiliary variables. The analogous generating function
for totally symmetric conserved currents built of complex  scalar fields has
been described  in  \cite{Bekaert:2009ud}. Introduce commuting auxiliary variables
$u^m u^n = u^n u^m$, $m,n = 0,...,d-1$ and define a bi-local function in $x$-variables and $u$-variables,
\be
\label{vienna1}
\cJ_{m_1... m_k}(x,u) = \bar \psi(x-u) \Gamma_{m_1... m_k} \psi(x+u)\;,
\qquad k = 0,..., d\;,
\ee
where spinor indices are implicit and $\Gamma_{m_1... m_k}$ are given by \eqref{scliff}. A conjugated
spinor is defined in such a way that the  contraction $\bar\psi^\alpha \psi_\alpha$
is invariant under Lorentz transformations, therefore  functions $\cJ_{m_1... m_k}(x,u)$ transform
as rank-$k$ antisymmetric Lorentz tensors. Then, introducing auxiliary anticommuting variables
$\theta^m \theta^n  = -\theta^n\theta^m$,  $m,n = 0,...,d-1$ one can define
extended generating function
\be
\label{vienna2}
\cJ(x,u|\theta) = \sum_{k=0}^d \cJ_{m_1... m_k}(x,u) \theta^{m_1} \cdots \theta^{m_k}\;,
\ee
such that functions \eqref{vienna1} appear as expansion coefficients with respect to
anticommuting $\theta$-variables.

Differentiating functions \eqref{vienna1} one obtains
\be
\label{vienna4}
\ba{l}
\dps
\frac{\d^2}{\d x^n \d u_n} \cJ_{m_1... m_k}(x,u) =
\\
\\
\dps
\hspace{20mm} =  - \Box \bar\psi(x-u) \Gamma_{m_1... m_k} \psi(x+u)
+\bar \psi(x-u) \Gamma_{m_1... m_k} \Box\psi(x+u)\approx 0\;,
\ea
\ee
where the weak equality symbol $\approx$ means that spinor fields satisfy massless Dirac equation,
and, as a consequence,  its quadrated version $\Box \psi_\alpha = 0$, $\Box \bar\psi^\alpha = 0$, where $\Box = \d^m \d_m$.
Another  differentiation yields the following expression
\be
\label{vienna3}
\ba{l}\dps
\frac{\d}{\d x_{n}} \cJ_{nm_1... m_k}(x,u) \approx
\\
\\
\dps
\hspace{20mm}\approx  - 2k\, \d_{[m_1} \bar \psi(x-u) \Gamma_{m_2... m_k]} \psi(x+u)
+2k\,  \bar \psi(x-u) \Gamma_{[m_1... m_{k-1}} \d_{m_k]}\psi(x+u)\;.
\ea
\ee
To prove the formula one uses
the gamma-matrix anticommutation relations and pulls
$\gamma_n$  to the
the left and the right sides of $\Gamma_{nm_1... m_k}$ in order to contract with $x$-derivatives and thereby produce respective  Dirac equations.

We conclude that the generating function \eqref{vienna2} fulfills the conditions
\be
\label{conserved}
\frac{\d^2}{\d x^m \d u_m} \cJ(x,u|\theta) \approx 0\;,
\qquad
\frac{\d^2}{\d x^m \d \theta_m} \cJ(x,u|\theta) \napprox 0\;.
\ee
The above expressions are reminiscent of current conservation conditions.
In the cases $k=0,1$ of  totally symmetric currents we see that the right-hand-side of \eqref{vienna3}
is weakly zero what precisely corresponds to a conservation condition. For $k\geq2$ the right-hand-side of \eqref{vienna3}
does not vanish.
However, an educated guess is that computing   derivatives and contracting indices  should be supplemented by proper
Young symmetrizations. In what follows, we  show that conservation conditions of one or another type
are valid for particular symmetry components contained in \eqref{vienna1}.

\subsection{ Flato-Fr\o nsdal conserved currents }
\label{sec:FF}

Expanding \eqref{vienna1} in power series of auxiliary
bosonic $u$-variable,
\be
\cJ_{m_1... m_p}(x,u) = \sum_{s=0}^\infty \frac{1}{s!}\, J_{a_1 ...a_s|m_1... m_p}(x) \, u^{a_1} ... u^{a_s}\;,
\ee
one obtains
\be
\label{current_basic}
J_{a_1 ...a_s|m_1... m_p}(x) = \sum_{t=0}^{s} (-)^{p+t} \d_{(a_1} ...  \d_{a_t} \bar \psi(x)
\Gamma_{m_1... m_p} \d_{a_{t+1}} ...  \d_{a_s)} \psi(x)\;.
\ee
Combinations  $J_{a_1 ...a_s|m_1... m_k}(x)$ are reducible tensors and their $gl(d)$
irreducible components are contained in a tensor product of
length $s$ symmetric row and height $k$ antisymmetric column. A list of irreducible
components is given by the following tensors: (I) type $\{s+1,p\}$ tensor,
(II) type $\{s,p+1\}$ tensor. Namely,
\footnote{We assume that a height $p$ does not exceed $[d/2]$. The case $p> [d/2]$ requires a separate
consideration that  presumably leads to dual current formulations.}
\be
\label{current1}
J^{(I)}_{a_1 ...a_{s+1}, \,m_1... m_{p-1}}(x)  = J_{(a_1 ...a_s|a_{s+1})m_1... m_{p-1}}(x)\;;
\ee
\be
\label{current2}
J^{(II)}_{a_1 ...a_{s}, \,m_1... m_{p}}(x)  = J_{a_1 ...a_{s}|m_1... m_{p}}(x)
- J_{(a_1 ...a_{s-1}[m_{1}|a_s)m_2... m_{p-1}]}(x)\;.
\ee

It is crucial that representatives of two families \eqref{current1} and \eqref{current2} with equal  spins $s$
have different  scale dimensions. Indeed, a spinor field in $\mathbb{R}^{d-1,1}$ has a scale dimension given by
\be
\label{spinorweight}
\Delta_0 = \frac{d-1}{2}\;,
\ee
while that of $\d_a$ is always $1$.
Then, scale dimensions of the same spin currents  $J^{(I)}_{a_1 ...a_{s}, \,m_1... m_{p}}(x)$
and $J^{(II)}_{a_1 ...a_s, \,m_1... m_p}(x)$ are given by \footnote{The other way around, a scale dimension of $J_{a_1 ...a_s|m_1... m_{k}}(x)$
equals $s+d-1$ and hence its two irreducible components \eqref{current1}, \eqref{current2} with spins shifted by one
 have equal dimensions.}
\be
\label{D1}
\Delta_1 = s+d-2\;,
\ee
\be
\label{D2}
\Delta_2 = s+d-1\;.
\ee
Our main conclusion here is that generating function $\cJ(x,u|\theta)$ \eqref{vienna2} yields
two families of mixed-symmetry currents with different scale dimensions. Currents
\eqref{current1} have dimensions coinciding with the unitarity
bound value $\Delta_1  = \Delta_{crit} $ \eqref{Dcrit}. Currents \eqref{current2} have bigger scale dimensions,
$\Delta_2 > \Delta_1$.

To deal with multi-index tensors  it is convenient to introduce the following differential  operators,
\be
\label{opers}
\ba{c}
\dps
S = \theta^m \frac{\d}{\d u^m}\;,
\qquad
S^* = u^m \frac{\d}{\d \theta^m}\;,
\\
\\
\dps
D = \frac{\d^2}{\d x^m \d u_m} \;,
\qquad
D^* = \frac{\d^2}{\d x^m \d \theta_m} \;.
\ea
\ee
These operators satisfy simple relations which  will be given  where appropriate. Then,
currents \eqref{current1} and \eqref{current2} can be represented as
\be
\label{FFcur}
\cJ^{FF}(x,u|\theta) = S^* \cJ(x,u|\theta)\;,
\ee
\be
\label{IMPcur}
\;\;\cJ^{imp}(x,u|\theta) = S^*S \cJ(x,u|\theta)\;,
\ee
where generating function $\cJ(x,u|\theta)$ is given by  \eqref{vienna2}.
By expanding these functions in auxiliary variables one obtains component forms \eqref{current1} and \eqref{current2},
respectively. Young symmetry conditions
 $S^* \cJ^{FF}(x,u|\theta) = 0$ and $S^* \cJ^{imp}(x,u|\theta) = 0$ are now obvious as a consequence  of   $S^* S^* = 0$.

Below  we formulate and prove three  Lemmas that elucidate the difference between  families of currents \eqref{FFcur}
and \eqref{IMPcur}.
We study whether or not currents are on-shell annihilated  by differential operators
$S^* S D$ and $D^*$ to be identified with two different types of conservation conditions discussed
in Section \bref{sec:genformcurrent}.

\begin{lemma}
\label{lemma1}
Currents \eqref{FFcur} satisfy the conditions
\be
\label{Conserv1}
S^* S D \cJ^{FF}(x,u|\theta) \approx 0\;,
\ee
\be
\label{popo}
D^* \cJ^{FF}(x,u|\theta) \napprox 0\;.
\ee
\end{lemma}
\noindent To  proof the lemma one uses relation \eqref{vienna4} represented as $D \cJ(x,u|\theta) \approx 0$,
and relation \eqref{vienna3} represented as
$D^* \cJ(x,u|\theta) \approx \alpha \,S \cJ(x,u|\theta)$, where $\alpha = \alpha(N_\theta)$ is some
function of the Euler operator counting a number of $\theta$-variables. Then, using
definition \eqref{FFcur}, (anti)commutation relations $S^* D^* + D^* S^*  =0$ and
$DS^* - S^*D = D^*$, one derives relations  \eqref{Conserv1} and \eqref{popo} by a direct computation.

Lemma \bref{lemma1} says that currents \eqref{FFcur}  satisfy the first type conservation condition \eqref{curcon}
only. Indeed, component form of  \eqref{Conserv1} reads
\be
\label{commmm}
\d^{n} J^{(I)}_{n a_2...a_{s-1}a_{s}, \,m_1m_2... m_{p-1}}(x)  -
\d^{n} J^{(I)}_{n (a_2...a_{s-1}[m_1,a_s) \,m_2... m_{p-1}]}(x) \,\approx\, 0\;,
\ee
and contracting above relation with auxiliary constant parameter
$\xi_1$ of $\{s-1,p\}$-type reproduces  \eqref{curcon}.
Another type of conservation law is not supported. Also, it is worth noting that the first type  conservation law is consistent with imposing
(anti-)selfdual conditions in even dimensions $d$, so currents with $p=d/2$ can be decomposed into their
(anti-)selfdual components.

Conservation  condition \eqref{commmm} trivializes for particular subclass of \eqref{FFcur}  identified
with totally antisymmetric Yukawa-like currents
\be
\label{Yukawa}
J^{(I)}_{m_1... m_{p}}(x) =  \bar \psi(x)\Gamma_{m_1 ... m_p} \psi(x)\;,
\qquad p \neq 1 \;,
\qquad
\Delta_1 = d-1\;,
\ee
that  generalize conventional Yukawa current $\bar \psi \psi$ and may be considered as
currents coupled to massive $k$-form fields. The case $p=1$ is exceptional, the corresponding current is the
well-known electromagnetic current $J^{(I)}_{m_1}(x) =  \bar \psi(x)\gamma_{m_1} \psi(x)$.

\begin{lemma}
\label{lemma3}
Currents \eqref{IMPcur} satisfy the conditions
\be
\label{Conserv12}
S^* S D \cJ^{imp}(x,u|\theta) \approx 0\;,
\qquad
D^*\cJ^{imp}(x,u|\theta) \approx 0\;.
\ee
\end{lemma}
\noindent The proof of  Lemma \bref{lemma3} goes along the same lines as the proof of Lemma \bref{lemma1}.
Note that by virtue of Young symmetry both conditions in \eqref{Conserv12} are algebraic consequences
of $D \cJ^{imp}(x,u|\theta) \approx 0$. Lemma \bref{lemma3} says that currents $\cJ^{imp}(x,u|\theta)$
satisfy conservation laws of both types.

\begin{lemma}
\label{lemma2}
Currents \eqref{FFcur} and \eqref{IMPcur} are related  as
\be
\label{relcu}
\cJ^{imp}(x,u|\theta) \approx D^* \cJ^{FF}(x,u|\theta)\;,
\ee
up to a normalization constant dependent on spins.
\end{lemma}
\noindent To  proof the lemma one uses relation \eqref{vienna3} represented as
$D^* \cJ(x,u|\theta) \approx \alpha \,S \cJ(x,u|\theta)$, where $\alpha = \alpha(N_\theta)$ is some
function of the Euler operator counting a number $\theta$-variables.  By a direct computation one obtains
$D^* \cJ^{FF}(x,u|\theta)\ = D^* S^* \cJ(x,u|\theta) = -S^* D^* \cJ(x,u|\theta)  = -\alpha_+ S^* S \cJ(x,u|\theta)
= -\alpha_+ \cJ^{imp}(x,u|\theta)$, where $\alpha_+ = \alpha(N_\theta+1)$; the last equality  reproduces definition \eqref{IMPcur}.
In particular, Lemma \bref{lemma2} justifies the reason why scale dimensions \eqref{D1} and \eqref{D2}
are shifted by one. \footnote{By  way of example consider current $J^{(II)}_a = \d_a \bar \psi \psi -
\bar \psi  \d_a \psi$ of dimension  $\Delta_2 = d$. The current  can be represented as a derivative of Yukawa-like
current $J^{(I)}_{[ab]}$ \eqref{Yukawa} of
dimension $\Delta_1 = d-1$ as $2J^{(II)}_a \approx \d^b J^{(I)}_{[ab]} \equiv \d^b(\bar \psi \Gamma_{ab}\psi)$. In particular,
the last equality demonstrates that Yukawa-like currents do not satisfy the conservation condition.}

Four  Lemmas \bref{lemma0} -  \bref{lemma2} taken together enable us to formulate the following propositions that play
the central role in this paper.
\begin{prop}
\label{prop1}
Currents \eqref{IMPcur} are improvements.
\end{prop}
\noindent To prove the proposition it is sufficient to note that according to Lemma \bref{lemma2} and Lemma \bref{lemma0}
conserved currents $\cJ^{FF}(x,u|\theta)$  are used to identify conserved currents $\cJ^{imp}(x,u|\theta)$
with trivial cohomology class .

\begin{prop}
\label{prop2}
Currents \eqref{FFcur} are Flato-Fr\o nsdal currents.
\end{prop}

\noindent The proposition is  obvious because  currents \eqref{FFcur} have scale dimensions and symmetry types that
exactly match  the spectrum of the generalized  Flato-Fr\o nsdal theorem \eqref{FF}, where the energies are equal to
scale dimensions $E_0 = \Delta_1  = s+d-2$ for massless fields (conserved currents),
and $E_0 = \Delta_1 = d-1$ for massive fields (Yukawa-like currents).

It is worth noting here that one-particle states of massless Dirac particle in $\mathbb{R}^{d-1,1}$  form
$o(d,2)$ module identified with singleton ${\rm Di}$. In particular, their quantum weights \eqref{singleton}
and \eqref{spinorweight} coincide.
It guarantees that all currents built of two spinors and finitely many
derivatives should have critical dimensions \eqref{Dcrit}, while any other possible conserved combinations
should come as \textit{on-shell}  improvements. Propositions \bref{prop1} and \bref{prop2} clearly confirm this observation.

In particular, it follows that  true conformal primary operators correspond to traceless conserved currents
of critical dimensions $\tilde{\cJ}^{FF}$ that can be constructed as $\tilde \cJ^{FF} = \cJ^{FF}+ ...$,
where the dots stand for the improvement terms.\footnote{Some examples of this procedure
for totally symmetric currents built of two spinors had been analyzed by Anselmi in \cite{Anselmi:1999bb}.} Note that  conservation and  tracelessness, as well as
adding improvements are to be  on-shell. \footnote{One can also add trivially conserved quantities proportional
to the projector matrix $\pi_{\mu\nu} = \d_\mu\d_\nu - \delta_{\mu\nu}\Box$ in such a way
that both conservation and traceless remain intact.} To support the above suggestion we mention
that any trace of the traceful conserved current $\cJ^{FF}$ is an on-shell improvement. Indeed, any trace
of the traceful conserved current is also conserved as can be seen
by taking traces on the conservation condition \eqref{commmm}. \footnote{This is true
for $s>2$ currents. For $s=2$ case the situation is different because the conservation condition
either has just one index ($p=1$), or has  antisymmetric indices only ($p\geq 2$). For spin-2
current the trace vanishes on-shell. For mixed-symmetry spin-$\{2,p\}$ currents the traces
are not conserved and do not vanish on-shell and  but all trace terms are organized into on-shell improvements, see next section.}
Then it follows that all traces are on-shell improvements because their spin numbers are smaller
than those of the original current, while their conformal dimensions are higher than corresponding
critical value.

\subsection{Traceless conserved currents: $s=2$, $p=2$ example}

\label{sec:exa}

As an example of the above discussion we consider the simplest case of three cell hook  conserved  current.
An explicit expression for the  traceful Flato-Fr\o nsdal  current read off from \eqref{FFcur}  is given by
(up to an overall factor)
\be
\label{hookcur}
\ba{r}
J_{a_1a_2, m_1}(x) =  - \d_{a_1}\bar\psi(x) \Gamma_{a_2 m_1}\psi(x) + \bar\psi(x) \Gamma_{a_2 m_1}\d_{a_1}\psi(x)
\\
\\
- \d_{a_2}\bar\psi(x) \Gamma_{a_1 m_1}\psi(x) + \bar\psi(x) \Gamma_{a_1 m_1}\d_{a_2}\psi(x) \;.
\ea
\ee
Its scale dimensions is $\Delta = d$ and Lemma \bref{lemma1} guarantees that this current is conserved,
\be
\d^{a_1}J_{a_1[a_2, m_1]}(x) \approx  0\;.
\ee
Traceless  component of \eqref{hookcur} denoted  by $\tilde J_{a_1a_2, m_1}$  is given by
\be
\label{trtr}
\tilde J_{a_1a_2, m_1}(x) = J_{a_1a_2, m_1}(x)
+ \frac{1}{2(d-1)} \big(\eta_{a_1m_1}J_{a_2}(x)+\eta_{a_2m_1}J_{a_1}(x) -2 \eta_{a_1a_2}J_{m_1}(x)\big)\;,
\ee
where $J_a(x) = \eta^{mn}J_{mn,a}(x)$. One observe that on-shell the trace is a total derivative
of the Yukawa current, $J_a(x) \approx 2\d_a(\bar\psi(x) \psi(x))$ and thereby the trace part in \eqref{trtr} is
an improvement. Indeed, going on-shell the trace part can be cast into the form $\d^n F_{a_1a_2,m_1n}(x)$, where
$F_{a_1a_2,m_1n}(x)  = \frac{1}{(d-1)} (\eta_{a_1m_1}\eta_{a_2n}+\eta_{a_2m_1}\eta_{a_1 n}
-2 \eta_{a_1a_2}\eta_{m_1n})J(x)$, where $J$ is the Yukawa current $J(x) = \bar\psi(x) \psi(x)$, while
the expression in the brackets has symmetries of the window Young diagram.
The conservation law is then easily obtained as
$\d^{a_1}\tilde J_{a_1[a_2, m_1]}(x) \approx -\frac{3}{2(d-1)} (\d_{a_2}J_{m_1}(x) - \d_{m_1}J_{a_2}(x)) \approx 0$.

A few comments are in order. First, it is important that proving conservation of the traceless current
is based  both on an algebraic subtraction of a trace from the traceful current, and  the on-shell realization of a trace as a divergence
of Yukawa current. Second, the above example can be directly generalized to arbitrary hooks of $\{2,p\}$-type;
here the trace is a derivative of the Yukawa-like current $\bar\psi(x)\Gamma_{m_1 ... m_{p-2}} \psi(x)$ so that
trace contributions are again on-shell improvements.
Third, it may also be worth noting that as the  current $J_{a_1a_2, m_1}(x)$ has only
three indices it follows that the double tracelessness condition is trivially satisfied anyway.

Finally, traceless mixed-symmetry conserved currents with arbitrary spins $s$ and $p$ can be constructed
by inserting some differential operator in between two spinors in the basic formula \eqref{vienna1}. This operation  should be
analogous to that one in the totally symmetric case  \cite{Giombi:2009wh,Bekaert:2009ud}. Note, however, that
choosing the traceless current cohomology representative is not unique and different choices are related to each
over by adding on-shell improvements, see \eqref{charge}, \eqref{closed}.

\section{Summary and discussion}
\label{sec:conclu}

In this paper we introduced conserved tensor currents of "hook" mixed-symmetry type in $d$-dimensional Minkowski space
and investigated  their basic properties. For special class of currents identified with $o(d,2)$
conformal primary fields   of critical dimensions we proposed an explicit realization via massless
spinor fields. On the other hand, we discussed primary fields of mixed-symmetry type for which
we classified conformally-invariant conditions with one derivative and described  two-point correlation functions.
Our  analysis of conserved currents supplemented by the generalized Flato-Fr\o nsdal theorem of \cite{Vasiliev:2004cm}
ensures the validity of the $AdS_{d+1}/CFT_d$ correspondence for "hook" mixed-symmetry fields on the kinematical level.
However, its dynamical realization may crucially depend on particular field-theoretical formulation used to describe bulk
fields \cite{Alkalaev_soon}.

It would be interesting to investigate the following applications of mixed-symmetry conserved currents.

First, an explicit realization of conserved currents via matter spinor fields  allows one to analyze
global symmetry algebra generated via conserved charges \eqref{charge}.
Indeed, constant $o(d-1,1)$ parameters of $\{s-1,p\}$ type
can be identified with a part of parameters of higher spin global symmetry algebra $hu(1|\,(1,2)\hspace{-1mm}:\hspace{-1mm}[d,2])$ introduced by Vasiliev
\cite{Vasiliev:2004cm}. To make the correspondence complete  one should proceed along the lines of
Ref.  \cite{Konstein:2000bi} and using Flato-Fr\o nsdal currents found in this paper construct
a full set of conserved currents with explicit $x$-dependence. It is expected that respective constant parameters are
described by $o(d,2)$ traceless tensors with symmetry type corresponding to two-row rectangular
block with one column added.

Also, once a higher spin theory in $AdS_{d+1}$ governed by $hu(1|\,(1,2)\hspace{-1mm}:\hspace{-1mm}[d,2])$ algebra is
formulated  it is  interesting to investigate its holographic dual in the spirit of the Klebanov-Polyakov and Sezgin-Sundell conjecture
in $d=3$ case \cite{Klebanov:2002ja,Giombi:2009wh}. It is tempting to speculate that
$hu(1|\,(1,2)\hspace{-1mm}:\hspace{-1mm}[d,2])$ higher spin theory is holographically
dual to the  Gross-Neveu model in $d$ dimensions. The mixed-symmetry Flato-Fronsdal currents built in this paper along with
the $AdS_4/CFT_3$ case provide a strong support in favor of such a duality. Of course, the conjecture crucially depends
on parity/higher spin symmetry violating and choosing either regular $\Delta_+ = d-1$ or irregular $\Delta_- = 1$
boundary conditions for massive antisymmetric bulk fields (including a scalar). All these important issues are to be clarified in as yet unknown higher spin theory for hook
fields.

One further direction concerns both  string theory and  higher spin theory interactions in Minkowski space. It would be interesting
to compute $n$-point correlation functions of mixed-symmetry
conserved currents along the lines of Refs.
\cite{Osborn:1993cr,Metsaev:2008fs,Sagnotti:2010at,Manvelyan:2010jr,Fotopoulos:2010ay,Giombi:2011rz,Costa:2011mg,Maldacena:2012sf,Stanev:2012nq}.

Finally, there remains an important task to describe  and give particular realizations of
conserved currents of arbitrary mixed-symmetry type in terms of dynamical fields.

\paragraph{Acknowledgements.} I am grateful  to M. Grigoriev, V. Didenko, R. Metsaev, E. Skvortsov  and M.A. Vasiliev for
useful   discussions and comments. I would like to thank the Erwin Schrodinger Institute, Vienna for hospitality
during my staying at ESI Workshop on Higher Spin Gravity, where the main part of this work has been done.  I am thankful
 to the Dynasty foundation for the financial support of the
visit. The work is supported in part by RFBR grant 11-01-00830 and RFBR grant 12-02-31838.

\providecommand{\href}[2]{#2}\begingroup\raggedright
\addtolength{\baselineskip}{-3pt} \addtolength{\parskip}{-1pt}

\providecommand{\href}[2]{#2}\begingroup\raggedright\endgroup

\end{document}